\documentclass[fleqn,10pt]{wlscirep}
\usepackage[utf8]{inputenc}
\usepackage[T1]{fontenc}

\usepackage{amsmath,amssymb,amsfonts}
\usepackage{epsfig}
\usepackage{epstopdf}
\usepackage{ifpdf} 
\usepackage{graphicx}
\usepackage{bm} 

\title{Noncentral forces mediated between inclusions in a bath of active Brownian rods}

\author[1,*]{Mahmoud Sebtosheikh}
\author[1,2,*]{Ali Naji}
\affil[1]{School of Physics, Institute for Research in Fundamental Sciences (IPM), P.O. Box 19395-5531, Tehran, Iran}
\affil[2]{School of Nano Science, Institute for Research in Fundamental Sciences (IPM), P.O. Box 19395-5531, Tehran, Iran}

\affil[*]{mahmoud-sebtosheikh@ipm.ir; a.naji@ipm.ir}

\begin{abstract}
Using Brownian Dynamics simulations, we study effective interactions mediated between two identical and impermeable disks (inclusions) immersed in a bath of identical, active (self-propelled),  Brownian rods in two spatial dimensions, by assuming that the self-propulsion axis of the rods may generally deviate from their longitudinal axis. When the self-propulsion is transverse (perpendicular to the rod axis), the accumulation of active rods around the inclusions is significantly enhanced, causing a more expansive steric layering (ring formation) of the rods around the inclusions, as compared with the reference case of longitudinally self-propelling rods. As a result, the transversally self-propelling rods also mediate a significantly longer ranged effective interaction between the inclusions. The bath-mediated interaction arises due to the overlaps between the active-rod rings formed around the inclusions, as they are brought into small separations. When the self-propulsion axis is tilted relative to the rod axis, we find an asymmetric imbalance of active-rod accumulation around the inclusion dimer. This leads to a noncentral interaction, featuring an anti-parallel pair of transverse force components and, hence, a bath-mediated torque on the dimer.  
\end{abstract}

\begin{document}

\flushbottom
\maketitle

\thispagestyle{empty}

\section{Introduction}
\label{sec:intro}

Active particles are characterized by their ability to take up ambient free energy and convert it to self-propelled motion \cite{Ramaswamy2010,revMarchetti2013}. They constitute a large class of particles, including molecular motors \cite{Bausch2010,Butt2010,Sumino2012,Inoue2015} that are powered by the hydrolysis of any available adenosine-tri-phosphate molecules, motile microorganisms such as bacteria and sperm cells \cite{Berg2003,Goldstein2015,Lardner1974,Woolley2003,Machemer1972,Sleigh1974} that self-propel via cellular organelles such as cilia and flagella, and artificial nano and microswimmers \cite{Paxton2004, Dreyfus2004,Howse2010,Sano2009,Gohy2017,Bechinger2012,Bechinger2016,Arslanova2021,Reddy2013,Dugyala2018,Vutukuri2016} such as Janus particles and functionalized colloids that undergo active motions due to catalyzing surface reactions or external stimuli such as light. The self-propelled motion drives any system of particles suspended in a base fluid out of thermodynamic equilibrium,  engendering a host of intriguing effects, including collective dynamic self-organization and  nonequilibrium pattern formation \cite{Ramaswamy2010,revMarchetti2013,revGompper2015,revLowen2016,revSpeck2020,revStark2016,revRomanczuk2012,revPeruani2019,Peruani2016,revVicsek2012}. 

Active particles are often subject to thermal (Brownian) and athermal (active) sources of noise, arising  typically from ambient fluctuations and internal mechanisms, respectively \cite{Ebeling2013,Schimansky-Geier2011,Romanczuk2013,Romanczuk2012,Schimansky-Geier2015}. Among various models of active motion \cite{revRomanczuk2012,revLowen2016,revPeruani2019}, active Brownian particles (ABPs)---featuring a constant self-propulsion speed subject to rotational Brownian noise---have emerged as a minimal model, capable of capturing key aspects of active particulate systems, with the greater part of  theoretical works on this model being devoted to computer simulations in two spatial dimensions \cite{revGompper2015,revLowen2016,revRomanczuk2012,revPeruani2019}. These include motility-induced phase separation (see, e.g., Refs. \cite{Baskaran2013,Cates2015,Cates2018} and references therein), nonequilibrium wall accumulation, boundary layering and barrier trapping  (see, e.g., Refs.\cite{Lowen2008,Gompper2009,Gompper2013,Gompper2015,Wensink2012,Wensink2013,Harder2014,Bolhuis2015,Ferreira2016,Naji2017,Naji2020z,Naji2020} and references therein), and collective self-organized motion  (see, e.g., Refs. \cite{Abkenar2013,Peruani2016,revPeruani2019,revVicsek2012} and references therein). Although these phenomena are enabled and sustained by particle activity, they can arise only in the presence of interparticle and particle-wall interactions, including Vicsek \cite{Vicsek1995,revPeruani2019,Peruani2016} and Vicsek-like \cite{Naji2021,Najafi2018,Najafi2017} alignment interactions and hydrodynamic interactions \cite{Najafi2018,Gompper2018}. Most of the aforementioned phenomena can emerge in active systems with only short-ranged  steric interactions between the constituent particles, signifying their generic nature. 

Models of ABPs with short-ranged steric interactions are particularly useful in the study of active colloidal suspensions in a base molecular liquid. When ABP dynamics is supplemented by the translational Brownian noise, the properties of the (bulk or confined) suspension can be tuned continuously from its reference equilibrium state, when particle self-propulsion is set to zero, to an arbitrary nonequilibrium steady state, merely by adjusting a single control parameter, which is known as P\'eclet number and is  defined as the ratio of the self-propulsion and rotational diffusivity timescales. The ABP framework thus facilitates direct comparisons between analogously designed active and passive systems. 

In equilibrium colloid science, bath-mediated effective interactions between colloidal inclusions have been of primary interest, given that such interactions directly determine the phase behavior of the suspended colloids. The inclusions can be of a whole range of shapes and dimensions (from nano to microscale), while the bath can, for instance, be a polar molecular solvent (as is often the case in the study of Casimir-van-der-Waals interactions) or a colloidal or polymeric suspension of smaller-size particles itself (as is often the case in the study of depletion forces) \cite{Israelachvili2011}. In the latter case, the effective bath-mediated interactions stem from the steric depletion of smaller bath particles (or depletants)  from the vicinity of the inclusions, yielding a short-ranged and attractive, depletion, interaction force between the inclusions  \cite{Lekkerkerker2011,Likos2001}.  

When colloidal inclusions are suspended in an active bath of ABPs \cite{revSpeck2020,Harder2014,Naji2017,Lowen2015,Cacciuto2014,Bolhuis2015,Leonardo2011,Reichhardt2014,Ferreira2016,Zhang2018,Zhang2020, Pagonabarraga2019,Kafri2018, Selinger2018, Garcia2015, Narayanan2018, Naji2020z, Yang2020, Marconi2018, Mishra2018, Naji2020}, the nature of bath-mediated interactions is altered significantly, as both attractive and repulsive forces of significantly larger  magnitude and range are possible, depending on the size and concentration of  the bath particles and, more importantly, also their activity strength. Active bath-mediated interactions between suspended inclusions can substantially be affected also by the presence of any confining boundaries \cite{Naji2020z}, shape, mobility and concentrations \cite{Yang2020,Mishra2018,Zhang2018,Naji2020z,Naji2020,Bolhuis2015} and, if present, dissimilarity \cite{Ferreira2016,Zhang2020} and permeability \cite{Naji2018,Naji2020} of the inclusions.  On the other hand, the shape of active particles itself plays a crucial role in determining their bulk properties  \cite{Baskaran2013,Orlandini2014,Filion 2016, Graaf 2016, Dunkel 2014,Gompper2013,Gompper2015}, near-wall accumulations at rigid  \cite{Lowen2008,Gompper2009,Gompper2013,Gompper2015,Naji2021} and deformable \cite{Angelan2016,Chen2017,Quillen2020} boundaries and, hence, the active (or `swim') pressure they exert on these boundaries \cite{Lowen2015,Kardar2015-1,Kardar2015-2,Naji2018,Naji2021}, and the effective interactions they mediate between external objects \cite{Leonardo2011,Harder2014,Zhang2018,Zhang2020}. For instance, the nonequilibrium wall accumulation of self-propelled rods turns out to be weaker than that of self-propelled disks \cite{Gompper2009,Gompper2013}. Active rods are often assumed to self-propel along their longitudinal (or long) axis and the torques experienced by the rods due to collisions with the wall reduces their near-wall `detention' time. The effects of active-particle shape on their clustering, segregation and collective motion have been explored for a range of different shapes such as active semiflexible filaments \cite{Glaser2021}, self-propelled triangles \cite{Graaf 2016}, squares \cite{Filion 2016}, crescents and other elongated different shapes with fore-aft asymmetry \cite{Dunkel 2014}. Self-propelled rods with longitudinal and transverse self-propulsion direction (being parallel and perpendicular to the longitudinal axis of the rods, respectively) have been realized experimentally  using, e.g.,  bimetallic Janus nano/microrods \cite{Paxton2004,Arslanova2021,Reddy2013,Dugyala2018,Vutukuri2016}. In particular, phase behaviors of transversally self-propelling rods have  numerically and experimentally been investigated \cite{Vutukuri2016}. Despite these recent works, systems of active rods with transverse  or, more generally, nonaxial (tilted) self-propulsion direction remain largely unexplored. 

Here, we focus on a minimal model of active Brownian rods with constant self-propulsion speed and, using  Brownian Dynamics simulations, explore the effects of nonaxial self-propulsion of the rods on their spatial distributions and the effective interactions they mediate between two colloidal inclusions.  The paper is organized as follows. We introduce our model and methods in Section \ref{sec:model} and discuss our simulation results for the distribution of the active Brownian rods in Section \ref{sec:distributions}, followed by an analysis of the bath-mediated interaction force  on  the inclusions in Section \ref{sec:eff_force}. The paper is summarized in Section \ref{sec:summary}.

\section{Model and Methods}
\label{sec:model}

\begin{figure}[t!]
\centering
\includegraphics[width=0.65\textwidth]{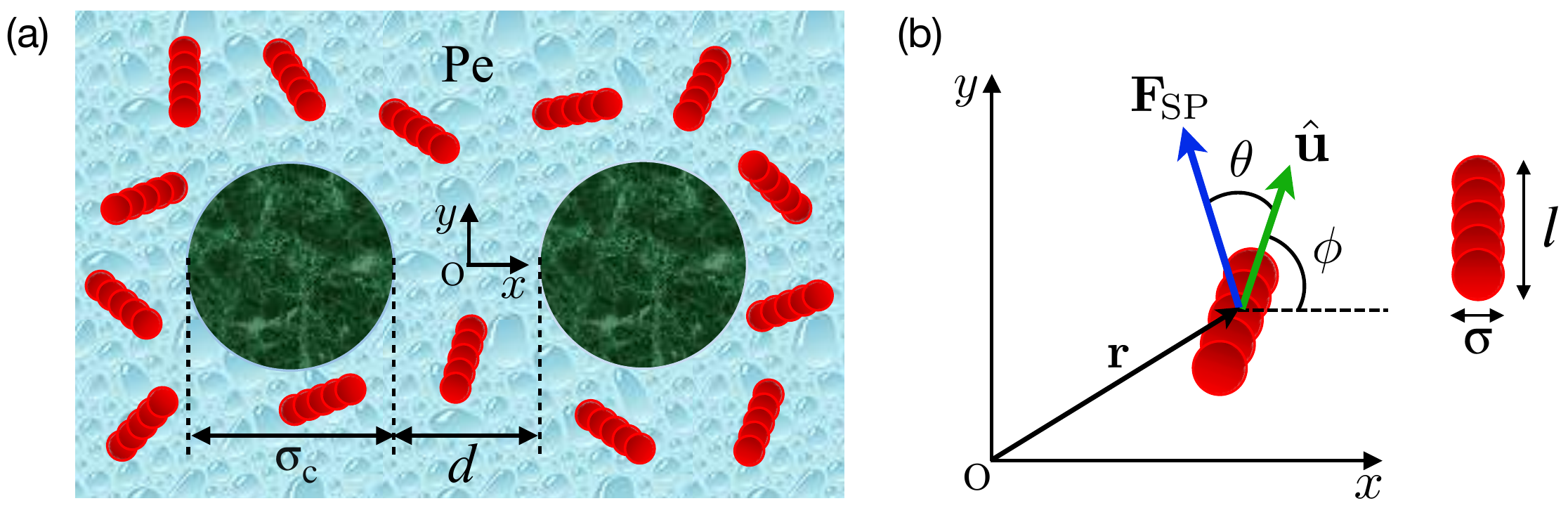}
\caption{(a) Schematic view of two fixed  inclusions of diameter $\sigma_\text{c}$ placed at intersurface distance $d$  in a bath of identical active Brownian rods. (b) Active rods are of width $\sigma$ and length $l$ and self-propel due to active forces $\mathbf{F}_\mathrm{SP}$ of fixed magnitude $F_\mathrm{SP}$ and self-propulsion angle $\theta$ relative to the orientation unit vector $\hat{\mathbf u}$ along the longitudinal axis of the rods. 
}
\label{fig:fig1}
\end{figure}

Our two-dimensional model comprises two identical, impermeable, disklike colloidal inclusions of diameter $\sigma_\text{c}$, placed at fixed positions at intersurface separation $d$ within a  bath of $N$ identical, stiff, active Brownian rods of width $\sigma$ and length $l$,  moving in a still liquid background of viscosity $\eta$; see Fig. \ref{fig:fig1}. The $x$-axis is taken along the center-to-center axis of the inclusions with the origin placed in the middle, yielding the center position vectors $\mathbf{R}_{1}=-\mathbf{R}_{2}=-(d+\sigma_\text{c})\hat{\mathbf x}/2$  for the `left' ($\mathbf{R}_{1}$) and the `right' ($\mathbf{R}_{2}$) inclusion in Fig. \ref{fig:fig1}a, with  $\hat{\mathbf x}$ being the $x$ unit vector. We consider two separate cases of interacting and noninteracting rods. In the former case, active rods repel one another via nearly hard and short-ranged, steric pair potentials, as will be specified later,   causing `contact' repulsive forces and torques between the rods. In the latter case, the active rods act as `phantom' particles with no interactions among themselves. In either case,  they always interact sterically with the disklike inclusions. 
 
The rods self-propel due to active forces $\mathbf{F}_{i,\mathrm{SP}}(t)$ of  fixed magnitude ${F}_\mathrm{SP}=|\mathbf{F}_{i,\mathrm{SP}}(t)|$ and fixed  {\em self-propulsion angle},  $\theta$, with $i=1,\cdots\,,\,N$ labeling the individual rods. The angle $\theta$ is defined relative to the instantaneous orientation unit vector $\hat{\mathbf u}_i(t)$; i.e., $\hat{\mathbf u}_i(t)\cdot \mathbf{F}_{i,\mathrm{SP}}(t)=F_\mathrm{SP}\cos\theta$. We define $\hat{\mathbf u}_i(t)=(\cos\phi_i(t),\sin\phi_i(t))$, where $\phi_i(t)$ is the parametrizing  polar angle measured from the $x$-axis; see Fig. \ref{fig:fig1}b. The configuration space of the active rods is thus spanned by  the center-of-mass position vectors $\{\mathbf{r}_i(t)\}=\{\left(x_i(t),y_i(t)\right)\}$ and the orientation vectors $\{\hat{\mathbf u}_i(t)\}$ that obey overdamped Langevin dynamics \cite{Doi-Edwards,Gompper-book}
\begin{eqnarray}
&&\dot{\mathbf{r}}_i={\mathbb M}_{\mathrm{T}}\cdot\left(-\nabla_{\mathbf{r}_i}U+ \mathbf{F} _{i,\mathrm{SP}}(t)+{\boldsymbol \xi}_{i,{\mathrm{T}}}(t)\right), 
\label{r}\\
&&\dot{\hat{\mathbf u}}_i=\left[{\mathbb M}_{\mathrm{R}}\cdot\left(-{\mathcal R}_{\hat{\mathbf u}_i}U+{\boldsymbol \xi}_{i,{\mathrm{R}}}(t)\right)\right]\times\hat{\mathbf u}_i(t), 
\label{theta}
\end{eqnarray}
where ${\mathcal R}_{\hat{\mathbf u}_i}={\hat{\mathbf u}_i}\times \nabla_{\hat{\mathbf u}_i}$ is the rotation operator \cite{Doi-Edwards,Gompper-book} and $\nabla_{\hat{\mathbf u}_i}$ is the unconstrained partial differentiation operator with respect to the Cartesian components  of  ${\hat{\mathbf u}_i}$. ${\mathbb M}_{\mathrm{T}}$ and ${\mathbb M}_{\mathrm{R}}$ are the relevant translational ($\mathrm{T}$)  and rotational ($\mathrm{R}$) mobility tensors,   
\begin{equation}
{\mathbb M}_{\mathrm{T}}=\zeta_{\parallel}^{-1}\hat{\mathbf u}_i\hat{\mathbf u}_i+\zeta_{\perp}^{-1}({\mathbb I}-\hat{\mathbf u}_i\hat{\mathbf u}_i), \quad 
{\mathbb M}_{\mathrm{R}}=\zeta_{\mathrm{R}}^{-1}{\mathbb I}, 
\label{RM}
\end{equation}
where ${\mathbb I}$ is the unit tensor, and $\zeta_{\parallel}$, $\zeta_{\perp}$ and $\zeta_{\mathrm{R}}$ are the (Stokes) friction coefficients associated with translation parallel ($\parallel$) and perpendicular ($\perp$) to the longitudinal axis of the rods and rotation ($\mathrm{R}$) in the $x-y$ plane, respectively. Also, in Eqs. \eqref{r} and \eqref{theta}, ${\boldsymbol \xi}_{i,{\mathrm{T}}}(t)=\xi_{i,\parallel}(t)\hat{\mathbf u}_i(t)+ \xi_{i,\perp}(t)\hat{\mathbf u}_i(t)\times\hat{\mathbf z}$ and ${\boldsymbol \xi}_{i,{\mathrm{R}}} (t)=\xi_{i,\mathrm{R}}(t)\,\hat{\mathbf z}$ are, respectively, the translational and rotational, Gaussian, white noises of zero mean, $\langle\xi_{i,\parallel}(t)\rangle=\langle\xi_{i,\perp}(t)\rangle=\langle\xi_{i,\mathrm{R}} (t)\rangle=0$, and two-point correlations $\langle\xi_{i,\parallel}(t)\xi_{j,\parallel}(s)\rangle = \zeta_{\parallel}k_\mathrm{B}T\delta_{ij}\delta(t-s)$,  $\langle\xi_{i,\perp}(t)\xi_{j,\perp}(s)\rangle = \zeta_{\perp}k_\mathrm{B}T\delta_{ij}\delta(t-s)$ and $\langle\xi_{i,\mathrm{R}}(t)\xi_{j,\mathrm{R}}(s)\rangle = \zeta_{\mathrm{R}}k_\mathrm{B}T\delta_{ij}\delta(t-s)$,  and $\hat{\mathbf z}$ is the out-of-plane (normal) unit vector. The friction coefficients depend on the length and aspect ratio  $p=l/\sigma$ of the rods and  the medium viscosity according to   \cite{Tirado1984} 
\begin{equation}
\!\!\zeta_{\parallel}=\frac{2\pi\eta l}{\ln p +\nu_{\parallel}},\,\,\, \zeta_{\perp}=\frac{4\pi\eta l}{\ln p +\nu_{\perp}},\,\,\, \zeta_{\mathrm{R}}=\frac{\pi\eta l^{3}}{3(\ln p +\nu_\mathrm{R})},
\label{zetas}
\end{equation}
where the so-called correction coefficients  $\nu_{\parallel}$, $\nu_{\perp}$ and $\nu_\mathrm{R}$ depend only on the aspect ratio and are given (for $2\leq p\leq 30$) by  \cite{Tirado1984}
\begin{eqnarray}
&&\nu_{\parallel}=-0.207 + 0.980/p - 0.133/p^2,\\
&&\nu_{\perp}=+0.839 + 0.185/p + 0.233/p^2,\\
&&\nu_\mathrm{R}=-0.662 + 0.917/p - 0.050/p^2.
\end{eqnarray}

Finally, in Eqs. \eqref{r} and \eqref{theta}, $U$ gives the sum of the interaction potentials between all `particles' in the system. We discretize each of the rods to a linear array of $n_\mathrm{b}=l/l_\mathrm{b}-1$ equidistant beads of diameter $\sigma$, with $l_\mathrm{b}=\sigma/2$ being the center-to-center distance between two adjacent beads in a given rod. In a given configuration of inclusions (labelled by $\alpha=1,2$) and active rods (labelled by $i, j=1,\cdots\,,\,N$, with their constituent beads labelled further by $a,b=1,\cdots\,,\,n_\mathrm{b}$), we have 
\begin{equation}
U=\sum_{i\neq j}\sum_{a,b} U_\mathrm{WCA}(\mathbf{r}_i^a,\mathbf{r}_{j}^b)+\sum_{i,a,\alpha}U_\mathrm{WCA}(\mathbf{r}_i^a,\mathbf{R}_{\alpha}), 
\end{equation}
where $\mathbf{r}_i^a$ is the position of the $a$th bead of the $i$th rod, etc, and $U_\mathrm{WCA}$ is  the Weeks-Chandler-Andersen (WCA) pair potential \cite{WCA1971}
\begin{equation}
U_\mathrm{WCA}({\mathbf r}, {\mathbf 0})=\left\lbrace 
\begin{array}{lll}
\!\!\!4\epsilon\left[\left(\frac{\sigma}{|{\mathbf r}|}\right)^{12}-2\left(\frac{\sigma}{|{\mathbf r}|}\right)^{6}+1\right]&:& |{\mathbf r}| \leq \sigma_\mathrm{eff},\\
\!\!\!0&:& |{\mathbf r}| >\sigma_\mathrm{eff}. 
\end{array}
\right.
\label{WCA}
\end{equation}
For the ensuing steric interaction between two individual beads (from two different rods), $|{\mathbf r}|$ is taken as  the center-to-center distance of the beads and $\sigma_\mathrm{eff}=\sigma$. For the interaction between a given bead and either of the inclusions, $|{\mathbf r}|$ gives the center-to-center bead-inclusion distance and $\sigma_\mathrm{eff}=(\sigma+\sigma_\text{c})/2$. In either case, we use the same interaction strength $\epsilon$. 

\subsection{Simulation methods and parameters}

We use standard Brownian dynamics methods \cite{McCammon1978} to numerically solve the governing equations \eqref{r} and \eqref{theta}. The relevant parameter space of the system can be spanned by the following set of dimensionless parameters: The rescaled inclusion diameter $\sigma_\text{c}/\sigma$, the rescaled intersurface distance of the inclusions, $d/\sigma$, the area fraction of active rods  $\Phi$, the self-propulsion angle $\theta$, and the {\em P\'eclet number} (relative strength of the self-propelling force),  $\mathrm{Pe}=\sigma F_\mathrm{SP}/(k_\mathrm{B}T)$.

Since our focus will be on the role of nonaxial self-propulsion, we fix the aspect ratio as $p=3$ ($n_\mathrm{b}=5$ beads), the inclusion diameter as $\sigma_\text{c}=12\sigma$ and the area fraction of active rods as $\Phi=0.2$ (see below). The P\'eclet number is varied over the interval $\mathrm{Pe}\in[5, 40]$, that covers a realistic range of values; e.g., synthetic active rods of length 2~$\mu \text{m}$ with self-propulsion speeds in the range 4-18~$\mu\text{m}/\text{s}$ have been used in experiments \cite{Paxton2004}, yielding P\'eclet numbers that, using our definitions, fall in the range $\text{Pe}\simeq 3-13$.  We restrict the self-propulsion angle to the first angular quadrant $\theta\in[0,\pi/2]$, as the results for other $\theta$-quadrants can be reproduced (within the simulation error bars) using appropriate   symmetry arguments; e.g., the results for a given self-propulsion angle, $\theta$, in the first quadrant can be mapped to those of the second quadrant angle $\pi-\theta$ using a left-right reflection (relative to the $y$-axis). For the sake of comparison with a monodisperse suspension of rods, we shall later consider two additional cases: A random suspension of active rods, whose self-propulsion angles are extracted from a uniform probability distribution over the interval $\theta\in[0,\pi/2]$, and a binary mixture of longitudinally ($\theta=0$) and transversally ($\theta=\pi/2$) self-propelling rods with relative fractions $1-n_{\pi/2}$ and  $0\leq n_{\pi/2}\leq 1$, respectively. 

In the simulations, we use a rectangular box of periodic boundaries of side lengths $L_x=L_y=53.5 \sigma$ and  $N=200$ active rods, giving the desirable area fraction $\Phi=\frac{N(p-1+\pi/4)\sigma^{2}}{(L^{2}-\pi\sigma_\text{c}^{2}/2)}\simeq 0.2$. 
We fix the WCA interaction strength as $\epsilon=20k_\mathrm{B}T$, ensuring sufficiently impermeable particles (note that strongly self-propelling rods can nevertheless partially penetrate into the inclusions, an effect that we shall return to in Section \ref{subsec:phantom}).  In the simulations, we use the timesteps of rescaled size $({3k_\mathrm{B}T}/{\pi\eta \sigma^{3}})\delta t=10^{-4}$ and run the simulations between   $1.3\times10^{7}$ and $5\times10^{7}$ timesteps, with $5\times 10^{6}$ initial steps used for relaxation to the desired steady state. The averaged quantities are then calculated by further averaging them over 36 statistically independent simulations. 

\begin{figure}[t!]
\centering
\includegraphics[width=\textwidth]{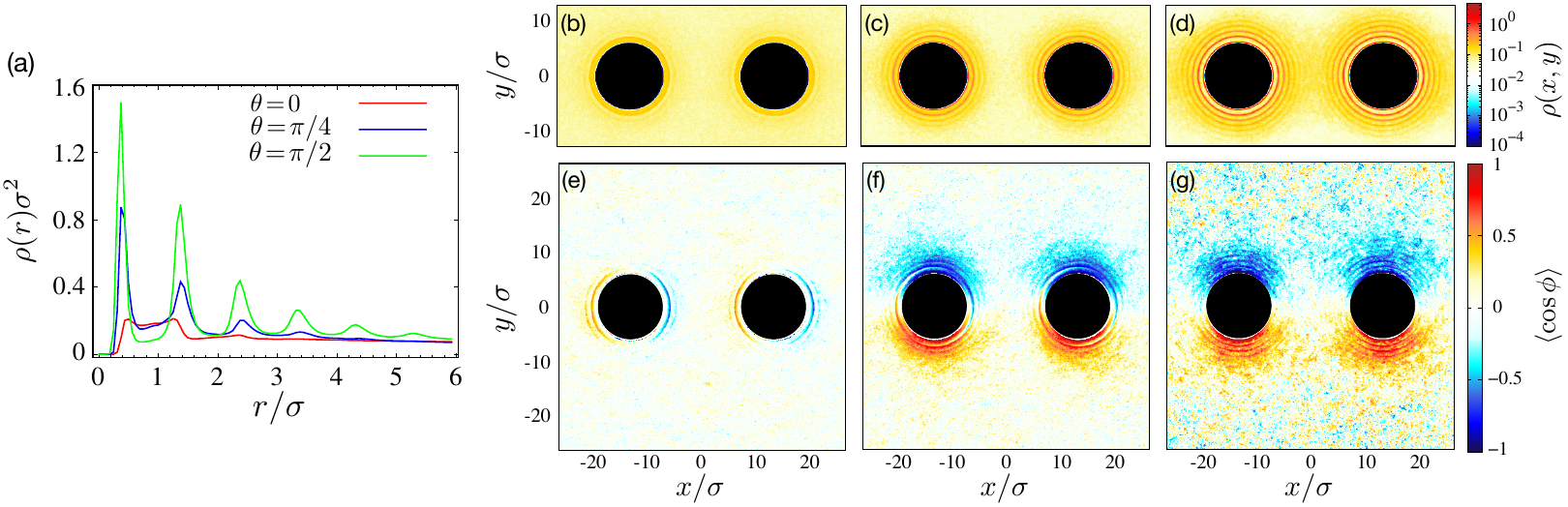}
	\caption{(a) Rescaled number density of active rods as a function of the rescaled radial distance, $r/\sigma$, from the surface of an isolated inclusion for  $\theta=0,\pi/4$ and $\pi/2$. Panels (b)-(d) show color-coded maps of the active-rod density field $\rho(x,y)$  for $\theta=0,\pi/4$ and $\pi/2$(left to right). Panels (e)-(g) show the corresponding  maps for  $\langle\cos \phi\rangle$. Here, $d/\sigma=14.5$ and $\mathrm{Pe}=20$. 
	}
\label{fig:fig2}
\end{figure}

\section{Results for spatial distribution of active rods}
\label{sec:distributions}

\subsection{Isolated inclusions}

We begin our discussion by examining the distributions of active rods around the disklike inclusions in the steady state. Figure \ref{fig:fig2}a shows the number density of active rods, $\rho(r)$, as a function of the radial distance, $r=|{\mathbf r}-{\mathbf R}_\alpha|-\sigma_c/2$, from the surface of an isolated inclusion (practically, either of the two inclusions $\alpha=1,2$ can be treated as an isolated one, provided that the inclusions are placed at sufficiently large intersurface distances to make them decoupled; here, we take  $d/\sigma=14.5$).  Being a well-known trait in systems of self-propelled particles \cite{Gompper2009,Gompper2013,Gompper2015}, the active rods in the present context also exhibit pronounced  accumulations in the close proximity of the inclusions. This is indicated by the near-surface peaks (occurring at spacings nearly equal to the rod width, $\sigma$) in the density profiles of Fig. \ref{fig:fig2}a. The peaks thus portray not only the excessive nonequilibrium accumulation of the rods near the inclusions, but also their steric layering, akin to that of disklike active particles, as extensively explored in Refs.\cite{Naji2017,Naji2020z,Naji2020}. In the absence of hydrodynamic effects and/or particle-specific features (such as swimmer tails), surface accumulation of active particles arises generically due to their prolonged detention near impermeable boundaries: Upon surface contact, the normal-to-surface component of the self-propulsion velocity pushes and holds the active particles against the surface, while its parallel-to-surface component persistently drives the particles along the boundary contour, until the rotational diffusion sets in, enabling particle escape from the boundary regions. In the case of active rods, this mechanism brings into play the nonaxial self-propulsion angle, $\theta$, as already indicated by the data in Fig. \ref{fig:fig2}a. 

The $\theta$-dependence is seen more clearly in the color-coded two-dimensional maps of the  density field,  $\rho(x,y)$, in the simulation box; see Figs. \ref{fig:fig2}b ($\theta=0$), c ($\pi/4$) and d ($\pi/2$). The density peaks of Fig. \ref{fig:fig2}a appear here as distinct high-density (dark yellow to red) layers, or `rings',  around the inclusions. As  $\theta$ increases from 0 to $\pi/2$ at fixed $\mathrm{Pe}$, and the longitudinal component of the self-propulsion velocity, $\mathrm{Pe}\cos \theta$, vanishes, the active rods show a significantly stronger surface accumulation (denser rings) over a wider range of radial distances from the inclusions (more rings). This picture can be   corroborated by $\langle\cos \phi\rangle$, which is plotted in Figs. \ref{fig:fig2}e ($\theta=0$), f ($\pi/4$) and g ($\pi/2$), giving a measure (first moment) of the rod orientation angle $\phi$ (the plotted quantity is defined as the local average of $\cos \phi$ {\em per particle} and, as such, the color intensity in the said panels should not be interpreted as the particle density). In regions away from the inclusions, the active rods adopt random orientations on average, yielding $\langle\cos \phi\rangle\simeq 0$. Close to the inclusions, the blue (red) arc segments with $  \langle\cos \phi\rangle\simeq 1$ (-1) highlight regions with typical rod axis orientations being nearly parallel (anti-parallel) to the $x$ axis, and white regions indicate rod axis orientations typically parallel or anti-parallel to the $y$ axis; in either case, the near-surface rod orientation is such that the self-propulsion direction of the rods points toward the inclusion interiors,  albeit not necessarily toward the inclusion centers. This can be seen for $\theta=\pi/4$ in panel f, where the centers of the blue/red arcs are found to be tilted from the $x$-axis not by $\pi/4$, but by an angle larger than $\pi/4$. For $\theta=0$ and $\pi/2$ (panels e and g), however, the near-surface rods typically orient their longitudinal axis along the local surface normal and tangent, respectively, and their self-propulsion directions points directly toward the inclusion centers. The tangent orientation of transversally self-propelling rods, pushing inwards from their sides edges, clearly signifies the dominant role of steric inter-rod and rod-inclusion interactions. This justifies the rods largest surface accumulation (longest near-surface detention) and most extensive ring formation around the inclusions, as seen in Figs. \ref{fig:fig2}a-d. 

\begin{figure}
\centering
\includegraphics[width=0.9\textwidth]{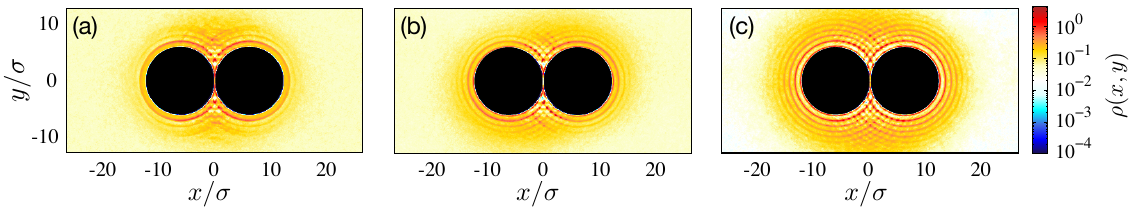}
	\caption{Panels (a), (b) and (c) show color-coded maps of the spatial distribution of active rods for $\theta=0,\pi/4$ and $\pi/2$, respectively with fixed  $\mathrm{Pe}=20$ and $d/\sigma=0.25$.  
	}
	\label{fig:fig3}
\end{figure}

\begin{figure}[t!]
    \centering
	\includegraphics[width=0.9\textwidth]{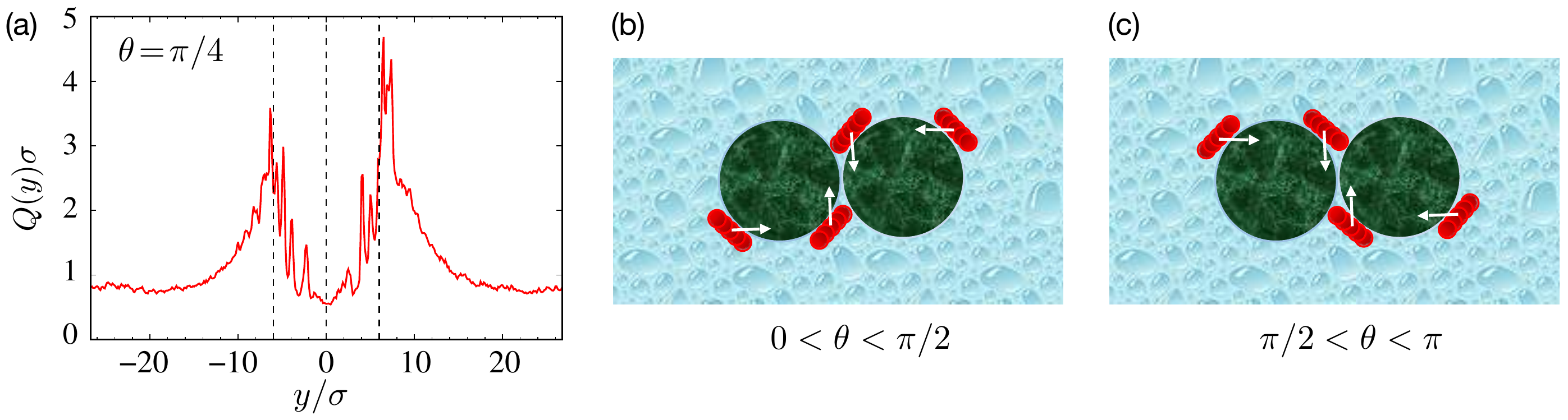}
	\caption{(a) Rescaled cumulative density of active rods $Q(y)\sigma$ defined along the $y$-axis and relative to the right inclusion (see the text for definitions) for $\text{Pe}=20$, $\theta=\pi/4$ and two juxtaposed inclusions at intersurface distance $d/\sigma=0.25$. The dashed lines indicate the extremities and the center level of the inclusion. Panels (b) and (c) show schematic views of the mechanism, causing asymmetric rod accumulation on the up versus down sides of the inclusions for $0<\theta<\pi/2$ and $\pi/2<\theta<\pi$.}
	\label{fig:fig4}
\end{figure}
 		
\subsection{Juxtaposed inclusions}
\label{subsec:juxtaposed}

When the two inclusions are brought into small intersurface separations, see Figs. \ref{fig:fig3}a-c with $d/\sigma=0.25$, the active rods are exposed to both types of convex and concave boundary curvatures. These types of curvatures occur, respectively, on the far sides of the inclusions and in the intervening region (the upper and lower wedge-shaped gaps) formed between the inclusions, respectively. In general, surface accumulation of active particles is stronger on concave boundaries as compared with convex ones \cite{Lowen2015,Gompper2015}. This effect enhances particle densities in the intervening gaps as opposed to those on the far sides of the inclusions. The relative accumulation in the gaps is strongest (weakest) in the case of longitudinally (transversally) self-propelling rods (panels a and c), making the spatial distribution of the rods around the {\em inclusion dimer} strongly (weakly) anisotropic. We shall discuss the ramifications of this strong versus weak anisotropy on the effective bath-mediated interactions in following sections.  

While the spatial distributions of active rods for  $\theta=0$ and $\pi/2$ (panels a and c) exhibit up-down symmetry, the situation turns out to be different for other values of $\theta$ (panel b for $\theta=\pi/4$). For the parameters in Fig. \ref{fig:fig3}, we find an {\em excess} ({\em deficit}) of active rods {\em above} ({\em below}) the {\em right} inclusion, with the situation being reversed in the case of the {\em left} inclusion. The asymmetric density imbalance can be seen more clearly in Fig. \ref{fig:fig4}a, where we plot the conventionally defined cumulative density,  $Q(y) = \int_{x_1}^{x_2}\rho(x,y)\,{\mathrm d}x$, of active rods in a vertical stripe $x\in[x_1,x_2]$, embedding the right inclusion, where $x_1=0$ and $x_2=\sigma_\text{c}+d/2+l$ (recall the inclusion diameter of $\sigma_\text{c}/\sigma=12$, the intersurface distance $d/\sigma=0.25$ and length of rod $l/\sigma=3$ in Fig. \ref{fig:fig3}, giving $x_2/\sigma=15.125$). The alternating peaks are again due to the active-rod ring formation around the inclusion, and the excess (deficit) of the rods above (below) the  inclusion are mirrored by the larger (smaller) peaks in the plot. 

We emphasize that the asymmetric up-down population imbalance emerges only when the inclusions are juxtaposed and when the self-propulsion axis is tilted $0<\theta<\pi/2$. The underlying mechanism for this effect can be understood by noting that, when the active rods come into contact with the inclusions, they adopt circular trajectories, going in counterclockwise direction around the inclusions. This type of chiral surface motion is driven by the longitudinal component of the self-propulsion, while the transverse component keeps the rods in contact with the inclusions. The effect is shown schematically in Fig. \ref{fig:fig4}b. For comparison, Fig. \ref{fig:fig4}c depicts the reversed clockwise motion that would arise for $\pi/2<\theta<\pi$. For the case considered here ($0<\theta<\pi/2$), the counterclockwise surface motion proceeds around the right (left) inclusion and is hindered, when the active rods reach the upper (lower) wedge-shaped gaps. Hence, the prolonged detention time and imbalanced crowding of the active rods above (below) the right (left) inclusion. 

\section{Results for effective interaction force}
\label{sec:eff_force}

Since the distribution of active particles around isolated inclusions is rotationally symmetric (Fig. \ref{fig:fig2}), the instantaneous forces imparted on the inclusions by the steric interactions (collisions) with the active particles average out to zero in the steady state. Any {\em net force} on the inclusions that may arise at small intersurface separations can thus be interpreted as the effective {\em interaction force} mediated by the active bath between the two inclusions. In the present context, such interactions originate from the steric overlaps between the active-particle rings formed around individual inclusions, as they are bought to small intersurface separations \cite{Naji2017,Naji2020z,Naji2020}. To proceed, we focus on the inclusion shown on the  {\em right} side ($\alpha=2$) in Fig. \ref{fig:fig1} and denote the effective force experienced by this inclusion by $\mathbf{F}(d)$. This quantity is computed in our simulations  using the relation
\begin{equation}
\label{F}
\mathbf{F}(d)=\sum_{i=1}^N\sum_{a=1}^n\bigg\langle\frac{\partial}{\partial {\mathbf{r}_i^a}}U_\mathrm{WCA}(\mathbf{r}_i^a,\mathbf{R}_{2})\!\bigg\rangle, 
\end{equation} 
where the brackets denote the steady-state average and the expression inside is the instantaneous forces imparted by the constituent beads of the rods on the said inclusion. $\mathbf{F}(d)=(F_x(d),F_y(d))$ depends on various system parameters, with the dependence on the intersurface distance, $d$, explicitly indicated. For brevity, we may refer to $\mathbf{F}(d)$ as the effective (interaction) force {\em between} or {\em acting on the inclusions} without further specifying it as the force on the right inclusion. One should bear in mind that the force on the left inclusion is of equal magnitude and opposite direction, as explicitly verified in the simulations (with the error bars). Thus, by definition, $F_x(d)>0$ ($F_x(d)<0$)  identifies a `repulsive' (`attractive') {\em longitudinal} force component, and $F_y(d)>0$ ($F_y(d)<0$) identifies an `upward' (`downward') {\em transverse} force component. When $F_y(d)\neq 0$, the effective interaction between the inclusions will be {\em noncentral}. In this case, there is an anti-parallel transverse force component of the same magnitude acting on the left inclusion and, hence, a {\em counterclockwise} ({\em clockwise}) torque on the inclusion dimer. 

\begin{figure}[t!]
    \centering
	\includegraphics[width=0.75\textwidth]{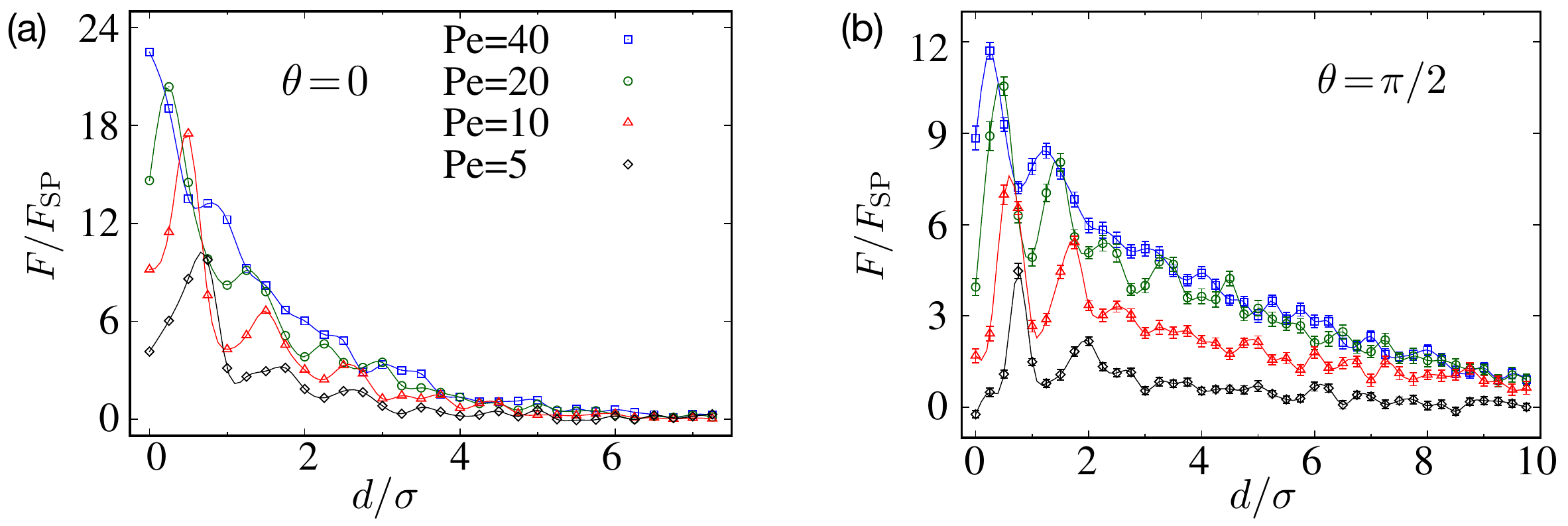}
	\caption{Rescaled effective force acting on the inclusions as a function of their rescaled intersurface distance, $d/\sigma$, for (a) longitudinally and (b) transversally self-propelling rods  and different values of $\mathrm{Pe}$, as indicated by the legends in panel a.}
	\label{fig:fig5}
\end{figure}

\subsection{Special cases with $\theta=0$ and $\pi/2$}
\label{subsec:0_pihalf}

Figure \ref{fig:fig5}a shows the simulated effective interaction force, Eq.   \eqref{F}, as a function of the intersurface distance, $d$, for different values of the P\'eclet number, $\mathrm{Pe}$, and in the special case of longitudinally (axially) self-propelling rods with $\theta=0$. The effective force in this case is found to be central; i.e., it  only has a longitudinal component, ${\mathbf F}(d)=F(d)\hat{\mathbf x}$. Its root cause is the broken left-right symmetry in the active-particle distribution around each individual inclusion. The preserved up-down symmetry in this case (Fig. \ref{fig:fig3}, panel a) ensures  $F_y=0$, as will  explicitly be verified later (see Fig. \ref{fig:fig6}b). 

The effective force $F(d)$ in Fig. \ref{fig:fig5}a is repulsive because of the previously discussed preferential accumulation of active rods in the wedge-shaped gaps between the inclusions (Section \ref{subsec:juxtaposed}). It exhibits a characteristic alternating profile  due to the successive intersections that occur between the active-particle rings of the two inclusions, as they are brought into surface contact (hence, the interpeak spacings in the force-distance profiles also exhibit a rather intricate pattern, even though they remains of the order of the rod width). These features of the force profiles in Fig. \ref{fig:fig5} are similar to those reported for active Brownian disks in Refs. \cite{Naji2017,Naji2020z,Naji2020}. The self-propelled rods, however,  experience inter-particle and particle-inclusion torques that exacerbate their trapping and accumulation in the wedge-shaped gaps between juxtaposed inclusions; an effect that is not  present in the case of disklike active particles. Increasing $\mathrm{Pe}$ enhances the magnitude of the effective force because of stronger rod-inclusion collisions, but suppresses its oscillatory behavior. The latter is due to  increased run lengths of active rods that, upon exceeding the inclusion size, lead to broadening and dilution of the circular rings they form around the inclusions.  

The force-distance profiles in the case of transversally self-propelling rods ($\theta=\pi/2$),  Fig.  \ref{fig:fig5}b, show similar features to those of longitudinally self-propelling rods. However, the effective force here is of relatively smaller  magnitude and of  larger range of action (compare Figs.  \ref{fig:fig5}a and b). The extended range of interaction can be understood by noting that the active-particle layering around the inclusions for $\theta=\pi/2$ appears over a respectively larger range of radial distances from each inclusion (Fig. \ref{fig:fig3}, compare panels a and c).  But since the distribution of transversally self-propelling rods around the inclusions is relatively more isotropic, as compared with the case of longitudinally self-propelling rods, the repulsive force due to the excess accumulation of active rods in the gaps between the inclusions is also relatively weaker.

\subsection{Tilted (nonaxial) self-propulsion}
\label{subsec:tilted}

\begin{figure}[t!]
\centering
	\includegraphics[width=0.75\textwidth]{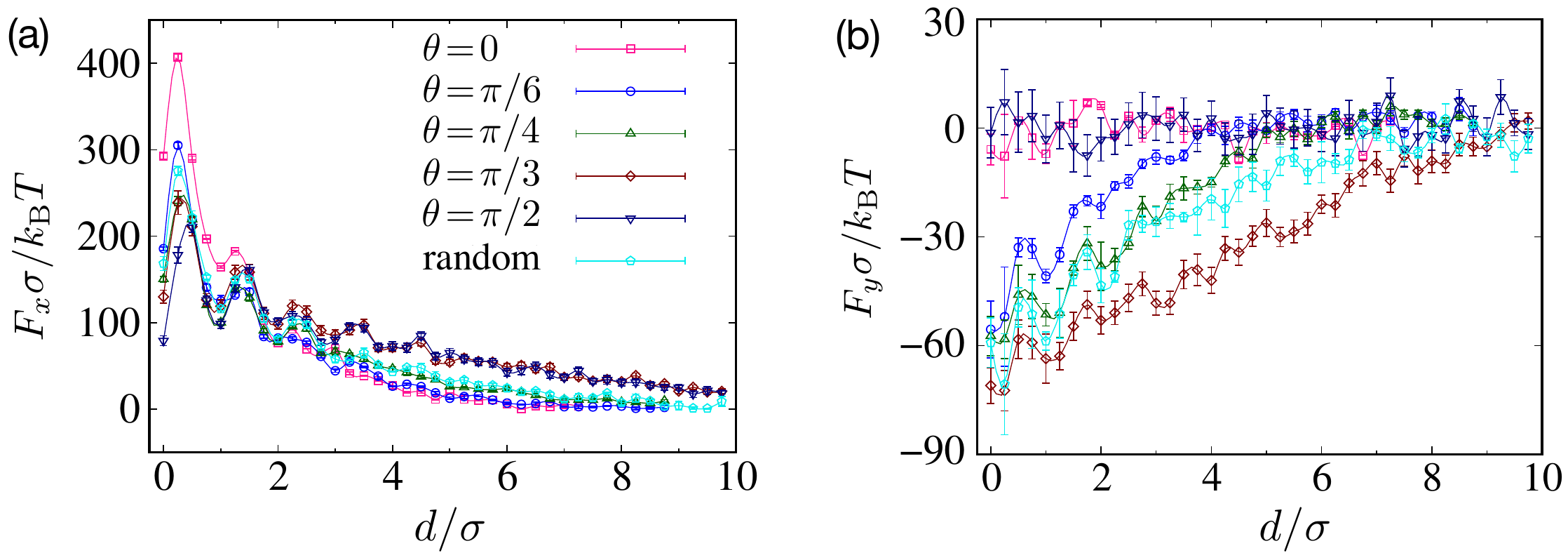}
	\caption{(a) Longitudinal ($F_x$) and (b) transverse  ($F_y$)  components of the rescaled effective force acting on the inclusions as a function of the rescaled intersurface distance, $d/\sigma$, for $\mathrm{Pe}=20$ and  different values of $\theta$, as indicated by the legends in panel a.}
	\label{fig:fig6}
\end{figure}

When the self-propulsion axis is tilted ($0<\theta<\pi/2$), the imbalance of active-rod population on the upper (lower) side of the right (left) inclusion, see Section \ref{subsec:juxtaposed}, yields  a nonvanishing transverse force component on the inclusions. In Figs. \ref{fig:fig6}a and b, we show the  longitudinal ($F_x$) and transverse ($F_y$) components of the effective force on the inclusions are shown as functions of the intersurface distance, $d$, for $\mathrm{Pe}=20$ and selected values of the self-propulsion angle $\theta$ (for the sake of demonstration, here we rescale the force components with $k_\mathrm{B}T/\sigma$). We also include the results for a random sample (cyan symbols) with $\theta$ in the active suspension being distributed uniformly within the interval $\theta\in [0,\pi/2]$. 

As seen in Fig. \ref{fig:fig6}a, at small separations $d/\sigma<2$, the peak amplitudes of the longitudinal interaction force, $F_x$, decrease as $\theta$ increases from 0 (red squares) to $\pi/2$ (dark blue triangle-downs); in other words, $F_x$ decreases as the self-propulsion direction becomes increasingly more tilted away from the rod axis. This is because, upon increasing $\theta$, active-rod accumulation around the inclusions becomes increasingly more isotropic (i.e., more active rods accumulate on the far sides of the inclusions), reducing the repulsive longitudinal force. The situation is {\em reversed} at large separations $d/\sigma>2$ and $F_x$ becomes consistently larger as $\theta$ increases from 0 to $\pi/2$; this is because $F_x$ falls off more weakly with the separation at larger $\theta$. The increase in the interaction range, as previously noted  for $\theta=\pi/2$ in Fig. \ref{fig:fig5}b, thus occurs continually as $\theta$ increases. For the most part, the data for the uniformly distributed (random) case fall closer to $\theta=\pi/4$ (green triangle-ups). Figure \ref{fig:fig6}b confirms that $F_y$ indeed vanishes  within the simulation error bars in the special cases of $\theta=0$ and $\pi/2$ over the whole range of intersurface separations. In other cases of $\theta$, including the case of randomly distributed $\theta$, $F_y$ is negative (i.e., according to our conventions, it pushes the left/right inclusion up/down, causing a  clockwise effective torque on the inclusion dimer) and shows  a distinctly long range. 

\begin{figure}[t!]
\centering
	\includegraphics[width=\textwidth]{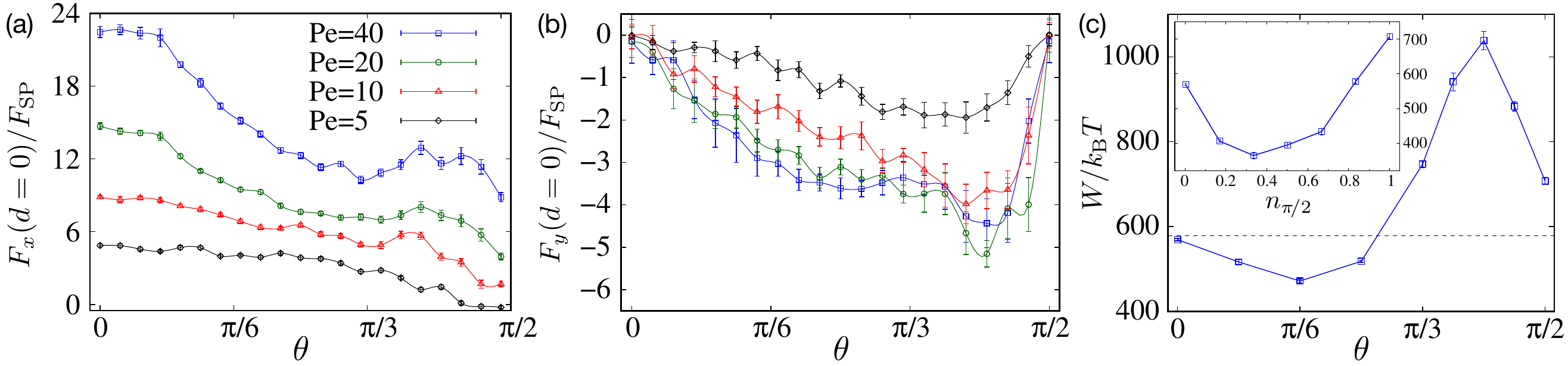}
	\caption{(a) Longitudinal ($F_x$) and (b) transverse  ($F_y$)  components of the rescaled effective force acting on the inclusions at surface contact $d=0$ as a function of $\theta$ for different values of $\mathrm{Pe}$, as indicated by the legends in panel a.  (c) The effective work, $W$, as defined in the text, plotted as a function of $\theta$ for $\mathrm{Pe}=20$. Symbols show  simulation data for a monodisperse suspension of active rods with given $\theta$, and the dashed line shows $W$ for a random dispersion with uniformly distributed $\theta\in[0,\pi/2]$. Inset shows $W$ for a binary mixture of longitudinally ($\theta=0$) and transversally ($\theta=\pi/2$) self-propelling rods as a function of the fraction of transversally self-propelling rods, $n_{\pi/2}$.
	}
	\label{fig:fig7}  
\end{figure}

Further insights into the $\theta$-dependent behavior of the interaction force can be obtained by fixing the intersurface distance as, e.g., $d=0$ (surface contact),  and examining the detailed behavior of the force components over the $\theta$ axis. The aforementioned trends in $F_x(d)$ are clearly mirrored by the longitudinal contact force, $F_x(d=0)$, shown in Fig. \ref{fig:fig7}a. As seen, $F_x(d=0)$ exhibits consistently larger (smaller) repulsive strengths at larger $\mathrm{Pe}$ (larger $\theta$). While, at small $\mathrm{Pe}$, we find a nearly monotonic decrease in $F_x(d=0)$ down to zero, the expected nonmonotonic alterations (due to stronger ring formations around the inclusions) emerge at large $\mathrm{Pe}$, with a pronounced hump in the subinterval $\pi/3<\theta<\pi/2$. The transverse contact force $F_y(d=0)$ in  Fig. \ref{fig:fig7}b expectedly vanishes at the end points of the depicted $\theta$-interval but the figure here reveals a pronounced global minimum (corresponding to a maximal force magnitude) in the subinterval $\pi/3<\theta<\pi/2$ at around $\theta \simeq 0.4\pi$. 

Another useful quantity that can complement the foregoing ones is the effective work required to bring the inclusions, in a quasistationary fashion from a sufficiently large intersurface distance, $d_{\infty}$, where the interaction force vanishes, into surface contact. We compute this quantity along the $x$-axis as $W=\int_{0}^{d_{\infty}}F_x(d)\,{\mathrm d}x$, by taking  $d_{\infty}/\sigma=10$ that gives vanishing interaction force within the simulation error bars. In an equilibrium setting with $\mathrm{Pe}=0$, $W$ corresponds to the so-called potential of mean force between the inclusions \cite{Lekkerkerker2011}. $W$ is shown in  Fig. \ref{fig:fig7}c, main set, as a function of the nonaxial self-propulsion angle, $\theta$. The symbols show the results for a monodisperse suspension of active rods with given $\theta$, while the dashed horizontal gives the corresponding value for a random dispersion of active rods, with uniformly distributed $\theta\in[0,\pi/2]$. As seen, for the parameters in the figure, $W$ in the case with $\theta=0$ (longitudinal self-propulsion) nearly matches that in the case of randomly distributed $\theta$. $W$ then drops from this value and decreases down to a global minimum at $\theta\simeq\pi/6$ and then turns up and matches the random sample (dashed line) at around $\theta\simeq0.27\pi$, growing to a global maximum at $\theta\simeq5\pi/12$, before falling off as $\theta$ tends to $\pi/2$. These  extrema signify the situations, where the overall repulsion between the inclusions is minimized/maximized. 

As defined, $W$ integrates the complex pattern of interaction force profiles (see Fig. \ref{fig:fig6}) into a single quantity. Even though the outcome (Fig. \ref{fig:fig7}c) gives a seemingly more tangible representation of the $\theta$-dependent behavior of the effective interactions, the detailed features of $W$ as a function $\theta$ (such as its limiting and extremal values)  remain to be understood. In the inset of Fig. \ref{fig:fig7}c, we show $W$ in the case of a binary mixture of {\em longitudinally} ($\theta=0$) and {\em transversally} ($\theta=\pi/2$) self-propelling rods with relative fractions $1-n_{\pi/2}$ and  $n_{\pi/2}$, respectively. $W$ is a convex function of $n_{\pi/2}$, maximized in the monodisperse cases ($n_{\pi/2}=0, 1$) and a globally minimized at $n_{\pi/2}\simeq 0.35$. This indicates that the overall repulsion between the inclusions can also be suppressed by producing a binary mixture of active rods. 

\subsection{Noninteracting active Brownian rods}
\label{subsec:phantom}

As a useful baseline to further elucidate the role of interparticle interactions in the present context, we consider an analogous system of noninteracting (`phantom') active Brownian rods by switching off (only) the steric inter-rod repulsions. Figures \ref{fig:fig8}a and b  show the corresponding interaction forces acting on the inclusions  for longitudinally ($\theta=0$) and transversally ($\theta=\pi/2$) self-propelling phantom rods, respectively. In these cases, the effective force is again central and  ${\mathbf F}(d)=F(d)\hat{\mathbf x}$. 

\subsubsection{Longitudinal self-propulsion ($\theta=0$)}

\begin{figure}[t!]
\centering
	\includegraphics[width=\textwidth]{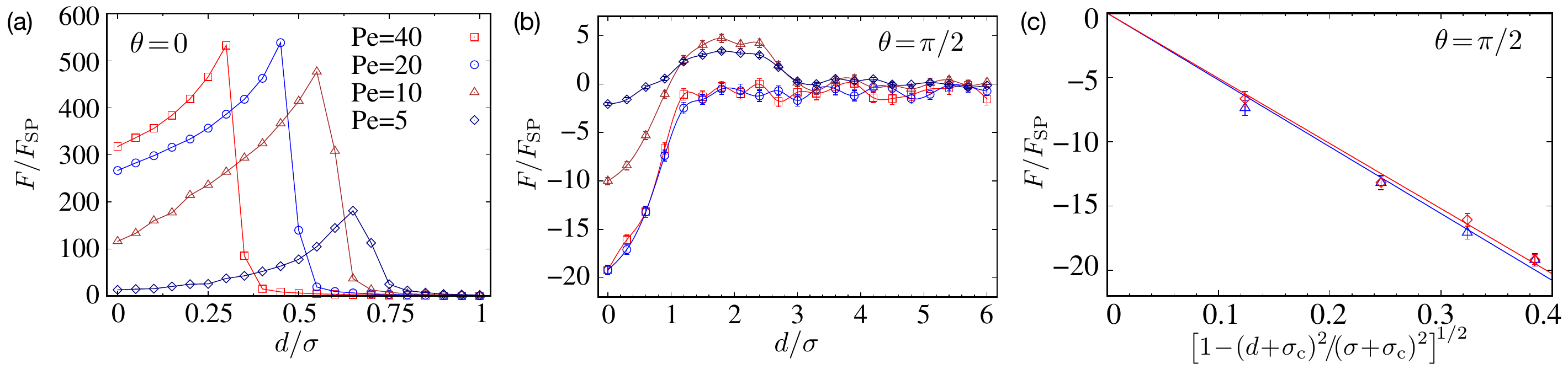}
	\caption{Rescaled effective force acting on the inclusions due to noninteracting (phantom) active rods as a function of the rescaled intersurface distance, $d/\sigma$, for (a) longitudinally and (b) transversally self-propelling rods  and different values of $\mathrm{Pe}$, as indicated by the legends in panel a. (c) Analytically predicted $d$-dependence of the active depletion-like force,  Eq. \eqref{eq:depletion}, is fitted  (lines) to the simulation data for $\theta=\pi/2$, $\text{Pe}=20$ and 40 (blue and red symbols, respectively) and $d<\sigma$. 
	}
	\label{fig:fig8}
\end{figure}

\begin{figure}[t!]
\centering
	\includegraphics[width=\textwidth]{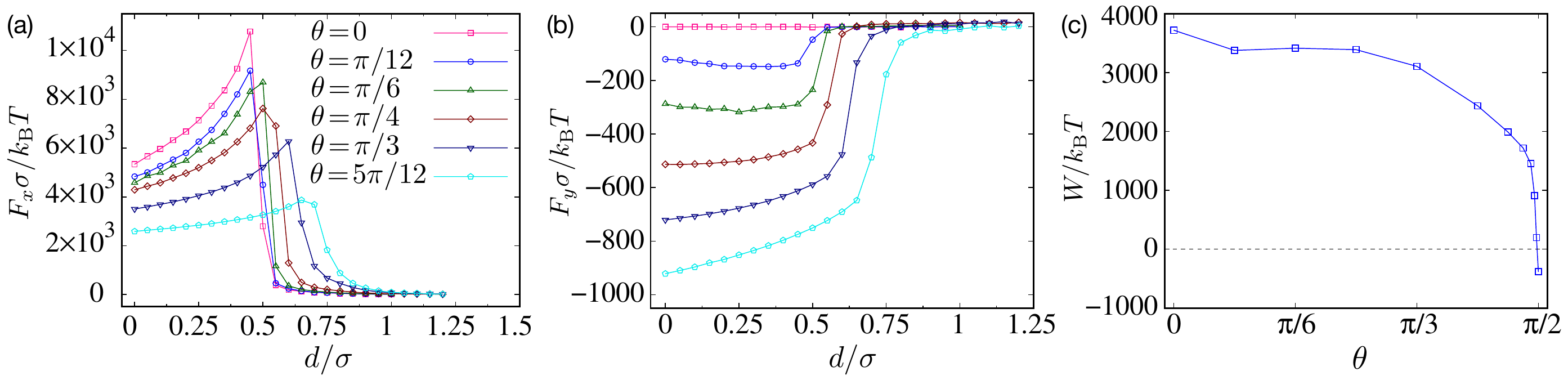}
	\caption{(a) Longitudinal ($F_x$) and (b) transverse  ($F_y$)  components of the rescaled effective force acting on the inclusions due to phantom active rods as a function of the rescaled intersurface distance, $d/\sigma$, for different values of $\theta$, as indicated by the legends in panel a. (c) Same as Fig. \ref{fig:fig7}c but here we plot the effective work, $W$, for phantom active rods. In all cases, $\mathrm{Pe}=20$.}
	\label{fig:fig9}
\end{figure}

For $\theta=0$, the effective force is repulsive, as in the interacting case, but it shows qualitatively different features: It is of significantly shorter range, of  significantly larger magnitude, and with only a single sharp peak, when compared with the case of interacting rods   (compare Fig. \ref{fig:fig8}a and Fig. \ref{fig:fig5}a). These differences reflect the differences that occur in the spatial distribution of active rods, when the steric inter-rod repulsions are turned off. For the sake of brevity (and given that the noninteracting model is an idealistic example and not at the focus of the present work), here we merely summarize some of the key aspects of the simulated phantom rod distributions, without any extensive presentation and discussion of the pertinent simulation data, but only as a means for elucidating the relevant interaction force profiles in Figs. \ref{fig:fig8} and \ref{fig:fig9}. 

The most notable difference between spatial distributions of phantom active rods and their interacting counterparts in Section \ref{sec:distributions} is that phantom rods do not form the  steric rings around the inclusions. Thus, in the case of longitudinally ($\theta=0$) self-propelling phantom rods, as the inclusions are brought into sufficiently small separations, the preferential accumulation of the rods in the concave gaps formed between the inclusions changes into their excessive overcrowding in those intervening regions. This leads to the much stronger effective repulsion between the inclusions with force strengths that can be larger by more than an order of magnitude (compare Fig. \ref{fig:fig8}a and Fig. \ref{fig:fig5}a), albeit in a significantly shorter range of separations, $d<\sigma$. The force-distance profiles in Fig. \ref{fig:fig8}a increase up to a maximum then suddenly drop to zero. The peak location indicates the typical gap width $d$ between the inclusions for which the active rods can start squeezing through the gap. This is enabled by partial penetration of the rods into the interfacial regions of the inclusions, as the rods self-propel from and to the top/bottom fluid regions connected by the intervening gap between the inclusions (note that the phantom rods can still interact sterically with the inclusions). These processes release the longitudinal stress and reduce the repulsive force, which appears as a sharp drop beyond the peak force. Hence, as the P\'eclet number $\mathrm{Pe}$ increases, the peak location in Fig. \ref{fig:fig8}a shifts to the left, as the stronger self-propulsion force helps the active rods to squeeze through even narrower intervening gaps (of widths down, e.g.,  to 45\% or 30\% of the rod width; see $\mathrm{Pe}=20$ and 40 in the figure). As $\mathrm{Pe}$ is increased  ($\text{Pe}\gtrsim 20$), the relative changes in the peak amplitude remains small, as total trapping of active rods in the intervening gap between the inclusions maintains the maximum longitudinal force acting on the inclusions at a fixed level. 
   
\subsubsection{Transverse self-propulsion ($\theta=\pi/2$): Depletion-like attraction}

For $\theta=\pi/2$, the absence of steric inter-rod repulsions and, primarily, the ring formation, of  phantom rods around the inclusions, also leads to significant modifications in the resulting interaction force as compared with the  interacting case (compare Fig. \ref{fig:fig8}b and Fig. \ref{fig:fig5}b). The features turn out to be quite different from those in the case of  $\theta=0$ as well (Fig. \ref{fig:fig8}a). The short-ranged effective force is found to be {\em attractive} in distances less than or comparable to the rod width, $d<\sigma$, followed by a weak repulsive hump, before leveling off to zero (Fig. \ref{fig:fig8}b). 

The effective force in Fig. \ref{fig:fig8}b resembles the typical force profiles in the context of equilibrium depletion interactions between colloidal particles \cite{Lekkerkerker2011,Likos2001}.  This finding  might appear unexpected, but it reflects the tendency of  transversally self-propelling phantom rods to adhere tangentially (from their side edges as pushed by the self-propulsion force) to the inclusions and to act similar to the so-called depletants in the said context. The key difference with the case of longitudinally self-propelling phantom rods is that--instead of overcrowding the intervening gap regions--transversally self-propelling phantom rods distribute rather isotropically around the circular boundaries of the inclusions, especially as the P\'eclet number, $\text{Pe}$, is sufficiently large (see below). Such a tendency was noted in the case of interacting rods as well (Section \ref{subsec:juxtaposed}), but the effect is exacerbated here by the absence of steric inter-rod repulsions between phantom rods. This leads to the formation of a thin interfacial layer of phantom rods around the inclusions with a thickness comparable to the rod width. When the inclusions are brought into small separations $d<\sigma$, the noted interfacial layer is sterically depleted from the intervening gap between the inclusions, engendering  a steric depletion mechanism similar to the traditional one in the equilibrium context \cite{Lekkerkerker2011,Likos2001} and, hence, a short-range attraction. 

 It is important to note that the resulting depletion-like attraction remains of predominantly active nature because of the finite P\'eclet number, and is thus of much greater magnitude than its equilibrium value ($\text{Pe}=0$). The attractive force behavior as a function of $d$ in Fig. \ref{fig:fig8}b can be estimated analytically, when $d$ is sufficiently small ($d<\sigma$). We assume  that $\text{Pe}$ is sufficiently large  ($\text{Pe}\gtrsim 20$) to warrant tangential and isotropic adherence of active rods around the inclusions (except in the intervening gap). The longitudinal force component due to a single rod of orientation angle $\phi$, adhered to the right inclusion and pushing against it, can be estimated as $\sim -F_\mathrm{SP}\sin\phi$, where $\phi$ is limited by the gap exclusion as $\phi\in[-\pi/2+\phi_d, 3\pi/2-\phi_d]$, with $\phi_d=\cos^{-1}\left[(d+\sigma_c)/(\sigma+\sigma_c)\right]$.  The longitudinal effective force is then estimated as
 \begin{equation}
F(d)\sim - \int_{-\pi/2+\phi_d}^{3\pi/2-\phi_d}F_\mathrm{SP}\sin\phi \,\text{d}\phi=-2F_\mathrm{SP}\left[1-(d+\sigma_c)^2/(\sigma+\sigma_c)^2\right]^{1/2},  
\label{eq:depletion}
\end{equation}
up to a prefactor that cannot be determined within the above analysis. The analytically predicted $d$-dependence of the active depletion-like force in Eq. \eqref{eq:depletion} fits the simulation data for $\text{Pe}=20$ and 40 very well (compare line and symbols in Fig. \ref{fig:fig8}c). 

The forgoing discussions elucidate the main qualitative and quantitative differences (in sign, magnitude and range) that we find between the effective forces mediated by longitudinally ($\theta=0$) and transversally ($\theta=\pi/2$) self-propelling phantom rods between the inclusions (Figs. \ref{fig:fig8}a and b). Our numerical inspections however indicate that the repulsive humps in the force profiles in Fig. \ref{fig:fig8}b arise as a result of yet another effect; namely, the transversally self-propelling phantom rods can get horizontally trapped in the wedge-shaped gaps between the inclusions, when the gap size is in the range  $\sigma<d<l$ (between the width of the rod and its length). This effect will be present only at weak to intermediate values of $\text{Pe}$, as otherwise the strong tendency of the rods to tangentially  adhere to the inclusions suppresses such rod-bridging configurations. 

\subsubsection{Tilted self-propulsion ($\theta=\pi/2$)}

Finally, for nonaxial self-propulsion angles $0<\theta<\pi/2$, the effective interaction between the inclusions becomes noncentral as in the case of interacting rods (Section \ref{subsec:tilted}). The $x$ and $y$-components of the resulting force, $F_x$ and $F_y$, as functions of the intersurface distance $d$ are shown in Figs. \ref{fig:fig9}a and b for $\mathrm{Pe}=20$ and a few different values of $\theta$. The longitudinal force profiles in Fig. \ref{fig:fig9}a display similar features to those with  $\theta=0$, even though the overall force magnitude decreases as $\theta$ increases. On close approach to purely transverse self-propulsion with $\theta=\pi/2$, the force magnitudes significantly reduce and the force profiles change to those of Fig. \ref{fig:fig8}b  (for illustration purposes, we have excluded $\theta=\pi/2$, whose data  become indiscernible in the scale of the graph). The transverse force component $F_y$, Fig. \ref{fig:fig9}b, again vanishes for the special cases $\theta=0$ and $\pi/2$; however, for other self-propulsions angles, the transverse force profiles due to phantom active rods also differ significantly from those  found in the case interacting rods (Fig. \ref{fig:fig6}b), even though $F_y$ is still negative and creates an effective clockwise torque on the inclusion dimer. The qualitative differences seen in the force profiles in the case of phantom rods as opposed to sterically interacting rods are reflected by the work, $W$, done on the inclusions as they are brought into contact along $x$-axis from far enough distances; see Fig.\ref{fig:fig9}c  (here, we take $d_\infty=6\sigma$). Unlike the case of interacting rods in Fig. \ref{fig:fig7}c, $W$ is convex here and varies mostly monotonically as a function of $\theta$, with a sharp drop as $\theta$ approaches $\pi/2$. The work drops negative values in this limit, signifying the attractive depletion-like force profiles of Fig. \ref{fig:fig8}b. 

\section{Summary}
\label{sec:summary}

Using extensive Brownian Dynamics simulations of a minimal model of active Brownian rods with  constant self-propulsion speed in two dimensions, we have investigated the role of nonaxial self-propulsion, identified by a tilt angle $\theta$, on (i) the steady-state distributions of the active rods around two fixed, impermeable, disklike inclusions immersed in the bath, and (ii) the effective interaction forces mediated by the active bath between the inclusions. Because the active rods are assumed to be rigid and interact with steric repulsions, they form layered structures (rings) around the inclusions at sufficiently high P\'eclet number, $\mathrm{Pe}$. This type of nonequilibrium particle layering occurs over a significantly larger range of radial distances from the inclusion boundaries in the case of transversally self-propelling rods ($\theta=\pi/2$) as compared with the case of longitudinally self-propelling rods ($\theta=0$). This in turn leads to effective interaction forces of respectively longer range of action between the inclusions in the former case. In both these cases, the interaction force is central, with only a longitudinal component along the center-to-center axis of the two inclusions. It is also repulsive because of a preferential accumulation of rods in the intervening wedge-shaped gaps formed between the inclusions at sufficiently small separations, even though the effect is weaker in the case of  transversally self-propelling rods, as they tend to adhere tangentially to the inclusions, displaying a relatively more isotropic spatial distribution around them. When the self-propulsion axis is tilted relative to the rod axis, the effective force acquires a finite transverse component and becomes noncentral. This effect is due an asymmetric imbalance of active-rod accumulation on the up and down sides of the inclusion dimer, and translates into a finite clockwise (counterclockwise) torque on the dimer for $0<\theta<\pi/2$ ($\pi/2<\theta<\pi$). For comparison purposes, we also presented simulations result for a random suspension of active rods with uniformly distributed self-propulsion angles, a binary mixture of longitudinally and transversally self-propelling rods, and the ideal case of phantom active rods, lacking steric inter-rod repulsions. 
  
Noncentral interactions have been reported in simulations of two hard inclusions embedded in a (dry) two-dimensional bed of granular disks,  activated by a shaking substrate with in-plane oscillations \cite{Pagonabarraga2019}. In this latter case, nonaxial self-propulsion is induced along a fixed external axis determined by the in-plane direction of substrate oscillations, contrasting the inherent self-propulsion tilt assigned to individual particles relative to their body axis in the present model. Our model is based on a first-step model of active rods in two dimensions. While (quasi-)two-dimensional realizations of the system would be feasible (with experimental examples ranging from  swarming bacteria \cite{Ariel2019} and shaken granular matter \cite{Tsimring2008} to   microswimmers confined to fluid-fluid interfaces \cite{Bishop2017} and/or confined by acoustic tweezers \cite{Brady2016}), it will be interesting to incorporate tractable features of real systems  within the present context as well. These include, e.g.,   hydrodynamic interactions between the constituent particles \cite{Najafi2018,Gompper2018},   possible confinement effects \cite{Naji2020z},   deformability  \cite{Glaser2021} and more complex shapes of self-propelled  particles  \cite{Dunkel 2014,Graaf 2016,Filion 2016}, and also mobility \cite{Yang2020,Mishra2018,Harder2014,Leonardo2011,Pagonabarraga2019},  shape dissimilarity  \cite{Ferreira2016,Zhang2020}  and permeability \cite{Naji2018,Naji2020} of the inclusions. 



\section*{Acknowledgements}
We thank the High Performance Computing Center (School of Computer Science, IPM) for computational resources and Salar Abbasi for useful comments. A.N. acknowledges the Associateship Scheme of The Abdus Salam International Centre for Theoretical Physics (Trieste, Italy).

\section*{Author contributions}
M.S. developed the numerical model/codes, performed the simulations, generated the output data and produced draft versions of the plots and the text. M.S. and A.N. conceived the study. A.N. supervised the research, produced the final figures and wrote the manuscript. Both authors analyzed the results, contributed to the discussions, and reviewed the text.

\section*{Competing interests}

The authors declare no competing interests.

\section*{Additional information}

{\bf Correspondence} and requests for materials should be addressed to M.S. or A.N.

\end{document}